\pdfoutput=1
\documentclass[12pt,a4paper]{article}
\usepackage{ifthen} % for conditional statements
\newboolean{pdflatex}
\setboolean{pdflatex}{true} % False for eps figures 

\newboolean{articletitles}
\setboolean{articletitles}{true} % False removes titles in references

\newboolean{uprightparticles}
\setboolean{uprightparticles}{false} %True for upright particle symbols

\newboolean{inbibliography}
\setboolean{inbibliography}{false} %True once you enter the bibliography

\usepackage[top=1in, bottom=1.25in, left=1in, right=1in]{geometry}

\usepackage{microtype}
\usepackage{lineno}  % for line numbering during review
\usepackage{xspace} % To avoid problems with missing or double spaces after
\usepackage{caption} %these three command get the figure and table captions automatically small

\usepackage{graphicx}  % to include figures (can also use other packages)
\usepackage{color}
\usepackage{colortbl}
\graphicspath{{./figs/}} % Make Latex search fig subdir for figures

\usepackage{amsmath} % Adds a large collection of math symbols
\usepackage{amssymb}
\usepackage{amsfonts}
\usepackage{upgreek} % Adds in support for greek letters in roman typeset

\newcommand*\patchAmsMathEnvironmentForLineno[1]{%
\expandafter\let\csname old#1\expandafter\endcsname\csname #1\endcsname
\expandafter\let\csname oldend#1\expandafter\endcsname\csname
end#1\endcsname
 \renewenvironment{#1}%
   {\linenomath\csname old#1\endcsname}%
   {\csname oldend#1\endcsname\endlinenomath}%
}
\newcommand*\patchBothAmsMathEnvironmentsForLineno[1]{%
  \patchAmsMathEnvironmentForLineno{#1}%
  \patchAmsMathEnvironmentForLineno{#1*}%
}
\AtBeginDocument{%
\patchBothAmsMathEnvironmentsForLineno{equation}%
\patchBothAmsMathEnvironmentsForLineno{align}%
\patchBothAmsMathEnvironmentsForLineno{flalign}%
\patchBothAmsMathEnvironmentsForLineno{alignat}%
\patchBothAmsMathEnvironmentsForLineno{gather}%
\patchBothAmsMathEnvironmentsForLineno{multline}%
\patchBothAmsMathEnvironmentsForLineno{eqnarray}%
}

\usepackage{hyperref}    % Hyperlinks in references
\usepackage[all]{hypcap} % Internal hyperlinks to floats.

\usepackage{xspace} 
\usepackage{upgreek}

\def\lhcb {\mbox{LHCb}\xspace}

\def\babar  {\mbox{BaBar}\xspace}
\def\belle  {\mbox{Belle}\xspace}

\def\MagUp {\mbox{\em Mag\kern -0.05em Up}\xspace}

\ifthenelse{\boolean{uprightparticles}}%
{

 \def\Pmu         {\ensuremath{\upmu}\xspace}                 
 \def\Pnu         {\ensuremath{\upnu}\xspace}                 
                  
 \def\Ppi         {\ensuremath{\uppi}\xspace}                 
                  
 \def\Prho        {\ensuremath{\uprho}\xspace}                 
                  
 \def\Ptau        {\ensuremath{\uptau}\xspace}

 \def\PDelta      {\ensuremath{\Delta}\xspace}                 
 \def\PXi      {\ensuremath{\Xi}\xspace}                 
 \def\PLambda      {\ensuremath{\Lambda}\xspace}                 
 \def\PSigma      {\ensuremath{\Sigma}\xspace}                 
 \def\POmega      {\ensuremath{\Omega}\xspace}                 
 \def\PUpsilon      {\ensuremath{\Upsilon}\xspace}                 
 
 \def\PB      {\ensuremath{\mathrm{B}}\xspace}                 
                  
 \def\PD      {\ensuremath{\mathrm{D}}\xspace}

 \def\PK      {\ensuremath{\mathrm{K}}\xspace}

 \def\Pb      {\ensuremath{\mathrm{b}}\xspace}

 \def\Pi      {\ensuremath{\mathrm{i}}\xspace}

 \def\Ps      {\ensuremath{\mathrm{s}}\xspace}

}
{

 \def\Pmu         {\ensuremath{\mu}\xspace}                 
 \def\Pnu         {\ensuremath{\nu}\xspace}                 
                  
 \def\Ppi         {\ensuremath{\pi}\xspace}                 
                  
 \def\Prho        {\ensuremath{\rho}\xspace}                 
                  
 \def\Ptau        {\ensuremath{\tau}\xspace}

 \mathchardef\PDelta="7101
 \mathchardef\PXi="7104
 \mathchardef\PLambda="7103
 \mathchardef\PSigma="7106
 \mathchardef\POmega="710A
 \mathchardef\PUpsilon="7107
                  
 \def\PB      {\ensuremath{B}\xspace}                 
                  
 \def\PD      {\ensuremath{D}\xspace}

 \def\PK      {\ensuremath{K}\xspace}

 \def\Pb      {\ensuremath{b}\xspace}

 \def\Pi      {\ensuremath{i}\xspace}

 \def\Ps      {\ensuremath{s}\xspace}

}

\makeatletter
\ifcase \@ptsize \relax% 10pt
  \newcommand{\miniscule}{\@setfontsize\miniscule{4}{5}}% \tiny: 5/6
\or% 11pt
  \newcommand{\miniscule}{\@setfontsize\miniscule{5}{6}}% \tiny: 6/7
\or% 12pt
  \newcommand{\miniscule}{\@setfontsize\miniscule{5}{6}}% \tiny: 6/7
\fi
\makeatother

\DeclareRobustCommand{\optbar}[1]{\shortstack{{\miniscule (\rule[.5ex]{1.25em}{.18mm})}
  \\ [-.7ex] $#1$}}

\def\mup        {{\ensuremath{\Pmu^+}}\xspace}
\def\mun        {{\ensuremath{\Pmu^-}}\xspace} % muon negative (\mum is taken)

\def\tauon      {{\ensuremath{\Ptau}}\xspace}
\def\taup       {{\ensuremath{\Ptau^+}}\xspace}
\def\taum       {{\ensuremath{\Ptau^-}}\xspace}

\def\neu        {{\ensuremath{\Pnu}}\xspace}
\def\neub       {{\ensuremath{\overline{\Pnu}}}\xspace}
\def\neum       {{\ensuremath{\neu_\mu}}\xspace}
\def\neumb      {{\ensuremath{\neub_\mu}}\xspace}
\def\neut       {{\ensuremath{\neu_\tau}}\xspace}
\def\neutb      {{\ensuremath{\neub_\tau}}\xspace}
\def\squark    {{\ensuremath{\Ps}}\xspace}

\def\bquark    {{\ensuremath{\Pb}}\xspace}

\def\pion   {{\ensuremath{\Ppi}}\xspace}
\def\piz    {{\ensuremath{\pion^0}}\xspace}

\def\pip    {{\ensuremath{\pion^+}}\xspace}
\def\pim    {{\ensuremath{\pion^-}}\xspace}

\def\rhomeson {{\ensuremath{\Prho}}\xspace}
\def\rhoz     {{\ensuremath{\rhomeson^0}}\xspace}

\def\kaon    {{\ensuremath{\PK}}\xspace}
  \def\Kbar    {{\kern 0.2em\overline{\kern -0.2em \PK}{}}\xspace}

\def\KorKbar    {\kern 0.18em\optbar{\kern -0.18em K}{}\xspace}

\def\Km      {{\ensuremath{\kaon^-}}\xspace}

  \def\Dbar    {{\kern 0.2em\overline{\kern -0.2em \PD}{}}\xspace}
\def\D       {{\ensuremath{\PD}}\xspace}

\def\DorDbar    {\kern 0.18em\optbar{\kern -0.18em D}{}\xspace}
\def\Dz      {{\ensuremath{\D^0}}\xspace}
\def\Dzb     {{\ensuremath{\Dbar{}^0}}\xspace}
\def\Dp      {{\ensuremath{\D^+}}\xspace}

\def\Dstar   {{\ensuremath{\D^*}}\xspace}

\def\Dstarm  {{\ensuremath{\D^{*-}}}\xspace}

\def\Dsp     {{\ensuremath{\D^+_\squark}}\xspace}

\def\Dssp    {{\ensuremath{\D^{*+}_\squark}}\xspace}

\def\B       {{\ensuremath{\PB}}\xspace}
\def\Bbar    {{\ensuremath{\kern 0.18em\overline{\kern -0.18em \PB}{}}}\xspace}

\def\BorBbar    {\kern 0.18em\optbar{\kern -0.18em B}{}\xspace}
\def\Bz      {{\ensuremath{\B^0}}\xspace}

\def\Bu      {{\ensuremath{\B^+}}\xspace}
\def\Bub     {{\ensuremath{\B^-}}\xspace}
\def\Bp      {{\ensuremath{\Bu}}\xspace}
\def\Bm      {{\ensuremath{\Bub}}\xspace}

\def\Bs      {{\ensuremath{\B^0_\squark}}\xspace}

  \def\Y#1S{\ensuremath{\PUpsilon{(#1S)}}\xspace}% no space before {...}!

\def\Lbar        {{\ensuremath{\kern 0.1em\overline{\kern -0.1em\PLambda}}}\xspace}
\def\LorLbar    {\kern 0.18em\optbar{\kern -0.18em \PLambda}{}\xspace}

\newcommand{\decay}[2]{\ensuremath{#1\!\to #2}\xspace}         % {\Pa}{\Pb \Pc}

\def\to                 {\ensuremath{\rightarrow}\xspace}

\def\AT#1     {\ensuremath{A_{\mathrm{T}}^{#1}}\xspace}           % 2

\def\C#1      {\ensuremath{\mathcal{C}_{#1}}\xspace}                       % 9
\def\Cp#1     {\ensuremath{\mathcal{C}_{#1}^{'}}\xspace}                    % 7
\def\Ceff#1   {\ensuremath{\mathcal{C}_{#1}^{\mathrm{(eff)}}}\xspace}        % 9  
\def\Cpeff#1  {\ensuremath{\mathcal{C}_{#1}^{'\mathrm{(eff)}}}\xspace}       % 7
\def\Ope#1    {\ensuremath{\mathcal{O}_{#1}}\xspace}                       % 2
\def\Opep#1   {\ensuremath{\mathcal{O}_{#1}^{'}}\xspace}                    % 7

\newcommand{\tev}{\ifthenelse{\boolean{inbibliography}}{\ensuremath{~T\kern -0.05em eV}}{\ensuremath{\mathrm{\,Te\kern -0.1em V}}}\xspace}
\newcommand{\gev}{\ensuremath{\mathrm{\,Ge\kern -0.1em V}}\xspace}
\newcommand{\mev}{\ensuremath{\mathrm{\,Me\kern -0.1em V}}\xspace}
\newcommand{\kev}{\ensuremath{\mathrm{\,ke\kern -0.1em V}}\xspace}
\newcommand{\ev}{\ensuremath{\mathrm{\,e\kern -0.1em V}}\xspace}
\newcommand{\gevc}{\ensuremath{{\mathrm{\,Ge\kern -0.1em V\!/}c}}\xspace}
\newcommand{\mevc}{\ensuremath{{\mathrm{\,Me\kern -0.1em V\!/}c}}\xspace}
\newcommand{\gevcc}{\ensuremath{{\mathrm{\,Ge\kern -0.1em V\!/}c^2}}\xspace}
\newcommand{\gevgevcccc}{\ensuremath{{\mathrm{\,Ge\kern -0.1em V^2\!/}c^4}}\xspace}
\newcommand{\mevcc}{\ensuremath{{\mathrm{\,Me\kern -0.1em V\!/}c^2}}\xspace}

\def\invfb   {\ensuremath{\mbox{\,fb}^{-1}}\xspace}

\newcommand{\stat}{\ensuremath{\mathrm{\,(stat)}}\xspace}
\newcommand{\syst}{\ensuremath{\mathrm{\,(syst)}}\xspace}

\newcommand{\chisq}{\ensuremath{\chi^2}\xspace}

\def\gsim{{~\raise.15em\hbox{$>$}\kern-.85em
          \lower.35em\hbox{$\sim$}~}\xspace}
\def\lsim{{~\raise.15em\hbox{$<$}\kern-.85em
          \lower.35em\hbox{$\sim$}~}\xspace}

\def\sqs   {\ensuremath{\protect\sqrt{s}}\xspace}

\def\pt         {\mbox{$p_{\mathrm{ T}}$}\xspace}

\def\evtgen     {\mbox{\textsc{EvtGen}}\xspace}

\def\geant      {\mbox{\textsc{Geant4}}\xspace}

\def\photos     {\mbox{\textsc{Photos}}\xspace}

\def\pythia     {\mbox{\textsc{Pythia}}\xspace}

\def\tell1  {TELL1\xspace}
\def\ukl1   {UKL1\xspace}

\usepackage{cite} % Allows for ranges in citations
\usepackage{mciteplus}

\usepackage{longtable} % only for template; not usually to be used in PAPERs

\newcommand{\extrn}{\ensuremath{\mathrm{\,(ext)}}\xspace}

\begin{document}

%%%%%%%%%%%%%%%%%%%%%%%%%
%%%%% Title     %%%%%%%%%
%%%%%%%%%%%%%%%%%%%%%%%%%
\renewcommand{\thefootnote}{\fnsymbol{footnote}}
\setcounter{footnote}{1}

% %%%%%%% CHOOSE TITLE PAGE--------
%\onecolumn
%\input{title-LHCb-INT}
%\input{title-LHCb-ANA}
%\input{title-LHCb-CONF}
% $Id: title-LHCb-PAPER.tex 95682 2016-07-21 12:13:58Z michaelt $
% ===============================================================================
% Purpose: LHCb-PAPER journal paper title page template
% Author: 
% Created on: 2010-09-25
% ===============================================================================

%%%%%%%%%%%%%%%%%%%%%%%%%
%%%%%  TITLE PAGE  %%%%%%
%%%%%%%%%%%%%%%%%%%%%%%%%
\begin{titlepage}
\pagenumbering{roman}

% Header ---------------------------------------------------
\vspace*{-1.5cm}
\centerline{\large EUROPEAN ORGANIZATION FOR NUCLEAR RESEARCH (CERN)}
\vspace*{1.5cm}
\noindent
\begin{tabular*}{\linewidth}{lc@{\extracolsep{\fill}}r@{\extracolsep{0pt}}}
\ifthenelse{\boolean{pdflatex}}% Logo format choice
{\vspace*{-2.7cm}\mbox{\!\!\!\includegraphics[width=.14\textwidth]{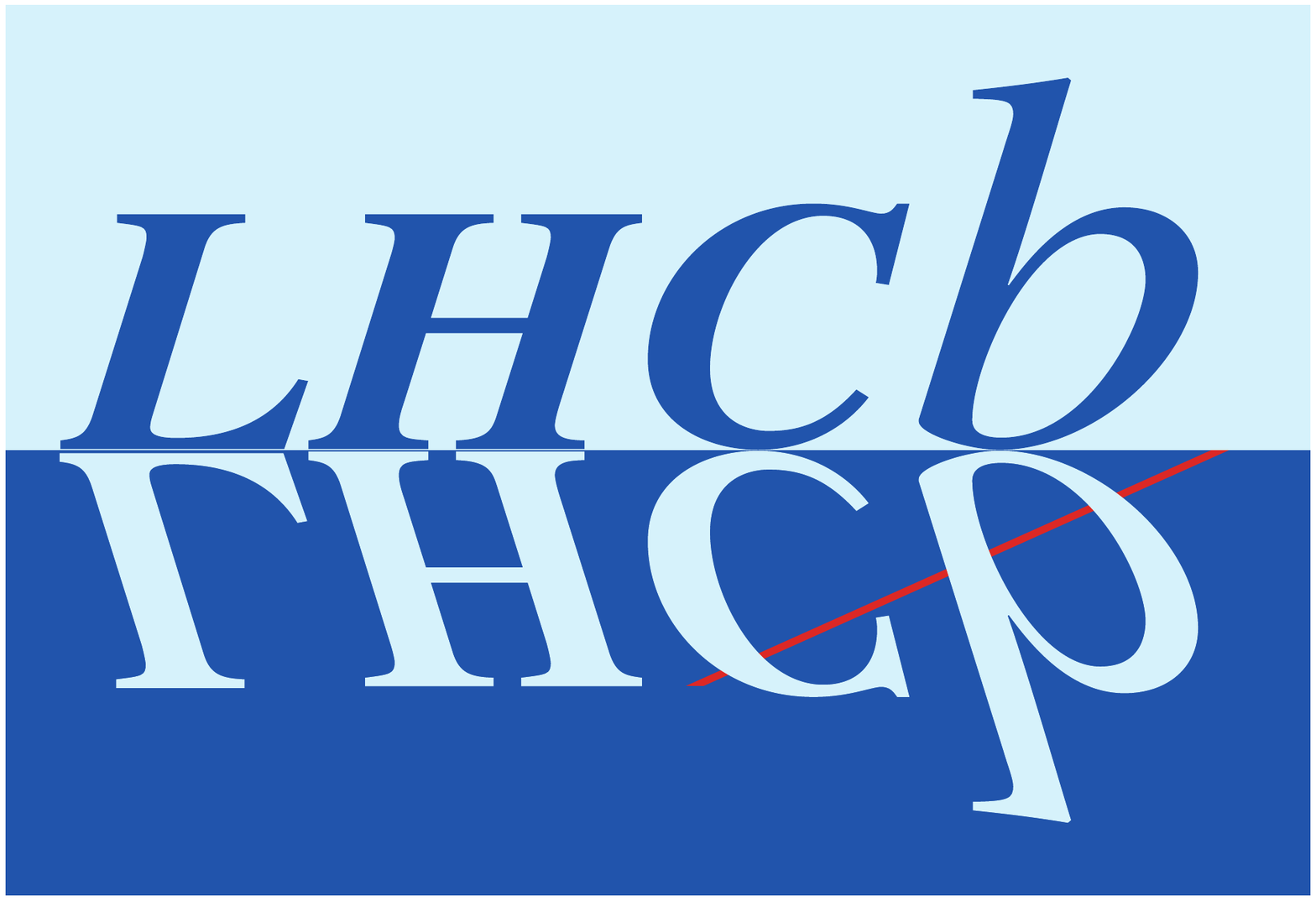}} & &}%
{\vspace*{-1.2cm}\mbox{\!\!\!\includegraphics[width=.12\textwidth]{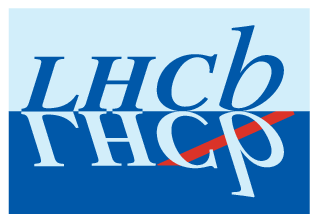}} & &}%
\\
 & & CERN-EP-2017-212 \\  % ID 
 & & LHCb-PAPER-2017-017 \\  % ID 
 & & 26 April 2018 \\ % Date - Can also hardwire e.g.: 23 March 2010
 & & \\
% not in paper \hline
\end{tabular*}

\vspace*{2.5cm}

% Title --------------------------------------------------
{\normalfont\bfseries\boldmath\LARGE
\begin{center}
      Measurement of %${\cal{B}}(B^0
      %\to D^{*-} \tau^+ \nu_{\tau})$ and 
      the ratio of the $B^0 \to D^{*-} \tau^+ \nu_{\tau}$ and $B^0 \to D^{*-} \mu^+
      \nu_{\mu}$ branching fractions %\mbox{${\cal{B}}(B^0
      %\to D^{*-} \tau^+ \nu_{\tau}) / {\cal{B}}(B^0 \to D^{*-} \mu^+
      %\nu_{\mu})$} \\
      using three-prong $\tau$-lepton decays
\end{center}
}

\vspace*{1.0cm}

% Authors -------------------------------------------------
\begin{center}
%In the footnote, replace 'paper' by 'Letter' in case of submission to PRL or PLB 
The LHCb collaboration\footnote{Authors are listed at the end of this paper.}
\end{center}

\vspace{\fill}

% Abstract -----------------------------------------------
\begin{abstract}
  \noindent
%    A measurement of %${\cal{B}}(B^0 \to D^{*-} \tau^+
    %\nu_{\tau})$ 
    The ratio of % semileptonic 
    branching fractions \mbox{${\cal{R}}(D^{*-})\equiv {\cal{B}}(B^0 \to D^{*-} \tau^+
    \nu_{\tau})/{\cal{B}}(B^0 \to D^{*-} \mu^+\nu_{\mu})$} is measured using a data
    sample of proton-proton collisions collected with 
    the LHCb detector  at center-of-mass energies of 7 and 8\tev, corresponding to an integrated
    luminosity of 3\invfb. 
 %HD   during the Run~1 of the LHC. 
For the first time ${\cal{R}}(D^{*-})$ is determined using the $\tau$
lepton decays  %HD through the hadronic
%HD    decay with 
with three charged pions in the final state. 
   %  Backgrounds from  prompt
   %  \B decays containing a $D^*$ and three pions are rejected by requiring that the $\tau$
   %  decay vertex lies downstream of the $B^0$ decay vertex. 
   % %
   %  Other physical backgrounds due to \B decays to double charmed
   %  hadrons are suppressed by partial reconstruction
   %  techniques, isolation criteria, and the dynamics of the 3$\pi$ system. A multi-dimensional fit is
   %  performed to extract the signal yield. 
 The $B^0 \to D^{*-} \tau^+\nu_{\tau}$ yield is normalized to that of the
 {\decay{\Bz}{\Dstarm\pip\pim\pip}} mode, providing a measurement of 
${\cal{B}}(B^0\to D^{*-}\tau^+\nu_{\tau})/{\cal{B}}(B^0\to
D^{*-}\pi^+\pi^-\pi^+) = 1.97 \pm 0.13 \pm 0.18$, where
the first uncertainty is statistical and the second systematic. The value of ${\cal{B}}(B^0 \to D^{*-} \tau^+
    \nu_{\tau}) = (1.42 \pm 0.094 \pm 0.129 \pm 0.054)\% $ is obtained, where the 
    third uncertainty is due to the limited knowledge of the branching
    fraction of the normalization mode. 
Using the well-measured branching fraction of the $B^0 \to
D^{*-} \mu^+\nu_{\mu}$ decay,  a value of  
${\cal{R}}(D^{*-}) = 0.291 \pm 0.019 \pm 0.026 \pm 0.013$ is
established, 
% . In both measurements, 
where the third uncertainty is due to the limited knowledge of the branching
    fractions of the normalization and 
    {\decay{\Bz}{\Dstarm\mup\nu_{\mu}}} modes. This measurement is in
    agreement with the Standard Model prediction and with previous
    results. 
\end{abstract}

\vspace*{1.0cm}

\begin{center}
  Published in Phys.~Rev.~Lett. {\bf{120}} (2018) 171802 
\end{center}

\vspace{\fill}

{\footnotesize 
\centerline{\copyright~CERN on behalf of the \lhcb collaboration, licence \href{http://creativecommons.org/licenses/by/4.0/}{CC-BY-4.0}.}}
\vspace*{2mm}

\end{titlepage}

%%%%%%%%%%%%%%%%%%%%%%%%%%%%%%%%
%%%%%  EOD OF TITLE PAGE  %%%%%%
%%%%%%%%%%%%%%%%%%%%%%%%%%%%%%%%

%  empty page follows the title page ----
\newpage
\setcounter{page}{2}
\mbox{~}
%\newpage
%
%% Author List ----------------------------
%%  You need to get a new author list!
%\input{LHCb_authorlist.tex}
%
%The author list for journal publications is provided by the Membership Committee shortly after 'approval to go to paper' has been given.
%%It will be made available on the page
%%\verb!http://www.physik.uzh.ch/~strauman/forMemCo/LHCb-PAPER-XXXX-XXX/! .
%It will be sent to you by email shortly after a paper number has beens assigned.
%The author list should be included already at first circulation, 
%to allow new members of the collaboration to verify whether they have been included correctly.
%Occasionally a misspelled name is corrected or associated institutions become full members.
%In that case, a new author list will be sent to you.
%In case line numbering doesn't work well after including the authorlist, try moving the \verb!\bigskip! after the last author to a separate line.
%
%
%The authorship for Conference Reports should be ``The LHCb
%  collaboration'', with a footnote giving the name(s) of the contact
%  author(s), but without the full list of collaboration names.

\cleardoublepage

%\twocolumn
% %%%%%%%%%%%%% ---------

\renewcommand{\thefootnote}{\arabic{footnote}}
\setcounter{footnote}{0}

%%%%%%%%%%%%%%%%%%%%%%%%%%%%%%%%
%%%%%  Table of Content   %%%%%%
%%%%%%%%%%%%%%%%%%%%%%%%%%%%%%%%
%%%% Uncomment next 2 lines if desired
%\tableofcontents
%\cleardoublepage

%%%%%%%%%%%%%%%%%%%%%%%%%
%%%%% Main text %%%%%%%%%
%%%%%%%%%%%%%%%%%%%%%%%%%

\pagestyle{plain} % restore page numbers for the main text
\setcounter{page}{1}
\pagenumbering{arabic}

%% Uncomment during review phase. 
%% Comment before a final submission.
%\linenumbers

In the Standard Model (SM) of particle physics, 
flavor-changing processes such as semileptonic decays of \bquark hadrons are mediated by a $W$ 
boson with universal coupling to leptons. 
Differences between the expected branching fraction of semileptonic
decays into the three lepton families originate from 
the different masses of the charged leptons. Lepton universality can be violated in
many extensions of the SM with nontrivial flavor
structure. 
Since uncertainties due to hadronic effects cancel to a large
extent, the SM prediction for the ratios between branching fractions of semitauonic
decays of \PB mesons relative to decays involving lighter lepton 
families, such as {\mbox{${\cal{R}}(D^{(*)-})\equiv {\cal{B}}(\Bz \to D^{(*)-} \taup
    \neut)/{\cal{B}}(\Bz \to D^{(*)-} \mup\neum)$}}
and  {\mbox{${\cal{R}}(D^{(*)0})\equiv {\cal{B}}(\Bm \to D^{(*)0} \taum
    \neutb)/{\cal{B}}(\Bm \to D^{(*)0} \mun\neumb)$}}, 
is known with an uncertainty at the percent
level~\cite{Fajfer:2012vx,Bigi:2016mdz,Bernlochner:2017jka,Jaiswal:2017rve}. The inclusion of charge-conjugate modes is implied throughout. 
These decays therefore provide a sensitive probe of SM extensions
with mass-dependent couplings, such as models with an extended Higgs sector~\cite{Tanaka:1994ay},
or leptoquarks~\cite{Buchmuller:1986zs,Davidson:1993qk}. 

Measurements of
${\cal{R}}(D^0)$, ${\cal{R}}(D^-)$, ${\cal{R}}(D^{*-})$, and ${\cal{R}}(D^{*0})$ have been reported by the
\babar~\cite{Lees:2012xj,Lees:2013uzd} and 
\belle~\cite{Huschle:2015rga,Sato:2016svk} collaborations in final states involving 
electrons or muons from the \tauon decay.  The \lhcb
collaboration published a determination of
${\cal{R}}(D^{*-})$~\cite{LHCb-PAPER-2015-025}, 
where the $\tau$ lepton was reconstructed using leptonic
decays to a muon.  The first simultaneous measurements of
${\cal{R}}(D^{*-})$, ${\cal{R}}(D^{*0})$, and $\tau$ polarization, using \tauon 
decays with one charged hadron in the final state, has recently  been published by
the \belle collaboration~\cite{Hirose:2016wfn}.  
All these  measurements yield values that are above the
SM predictions with a combined significance of
3.9 standard deviations~\cite{HFAG}. 

This Letter reports the first determination of ${\cal{R}}(\Dstarm)$
using the three-prong $\taup\to\pip\pim\pip\neutb$ and
$\taup\to\pip\pim\pip\piz\neutb$ decays. 
 A more detailed description
of this measurement is given in Ref.~\cite{LHCb-PAPER-2017-027}. 
The \Dstarm meson is reconstructed through the
\decay{\Dstarm}{\Dzb (\to K^+ \pim) \pim} decay chain. The visible final
state consists of six charged tracks; neutral pions are
ignored in this analysis. A data sample of proton-proton
collisions, corresponding to an integrated luminosity of 3\invfb, 
collected with the \lhcb detector at center-of-mass energies \sqs=
7 and 8\tev is used. 

In order to
reduce experimental systematic uncertainties, the $B^0\to
D^{*-}\pi^+\pi^-\pi^+$ decay is chosen as a normalization
channel. This 
leads to a measurement of the ratio  
\begin{equation}
{\cal{K}}(\Dstarm)\equiv\frac{{\cal{B}}(B^0 \to D^{*-} \tau^+
    \nu_{\tau})
}{{\cal{B}}(B^0 \to
    D^{*-} 3\pion)} =
    \frac{N_{\mathrm{sig}}}{N_{\mathrm{norm}}}\frac{\varepsilon_{\mathrm{norm}}}{\varepsilon_{\mathrm{sig}}}\frac{1}{{\cal{B}}(\tau^+\to
    3\pi\neutb)+{\cal{B}}(\tau^+\to 3\pi\piz\neutb)},
\label{eqn:kappa}
\end{equation}
where $3\pion\equiv\pip\pim\pip$, and $N_{\mathrm{sig}}$ ($N_{\mathrm{norm}}$) and $\varepsilon_{\mathrm{sig}}$
($\varepsilon_{\mathrm{norm}}$) are the yield and selection efficiency
for the signal (normalization) channel, respectively. From this, ${\cal{R}}(D^{*-})$
is obtained as ${\cal{R}}(D^{*-}) = {\cal{K}}(\Dstarm) \times {\cal{B}}(B^0 \to
    D^{*-} 3\pion) / {\cal{B}}(B^0 \to
    D^{*-} \mup\neum)$, where the branching fraction of the 
    \decay{\Bz}{\Dstarm3\pion} decay is taken as the weighted average of the
    measurements of
    Refs.~\cite{LHCb-PAPER-2012-046,TheBABAR:2016vzj,Majumder:2004su},
    and that of the \decay{\Bz}{\Dstarm\mup\neum} decay 
is taken from Ref.~\cite{HFAG}.   

\begin{figure}[t]
	\centering
	\includegraphics[width=0.6\textwidth]{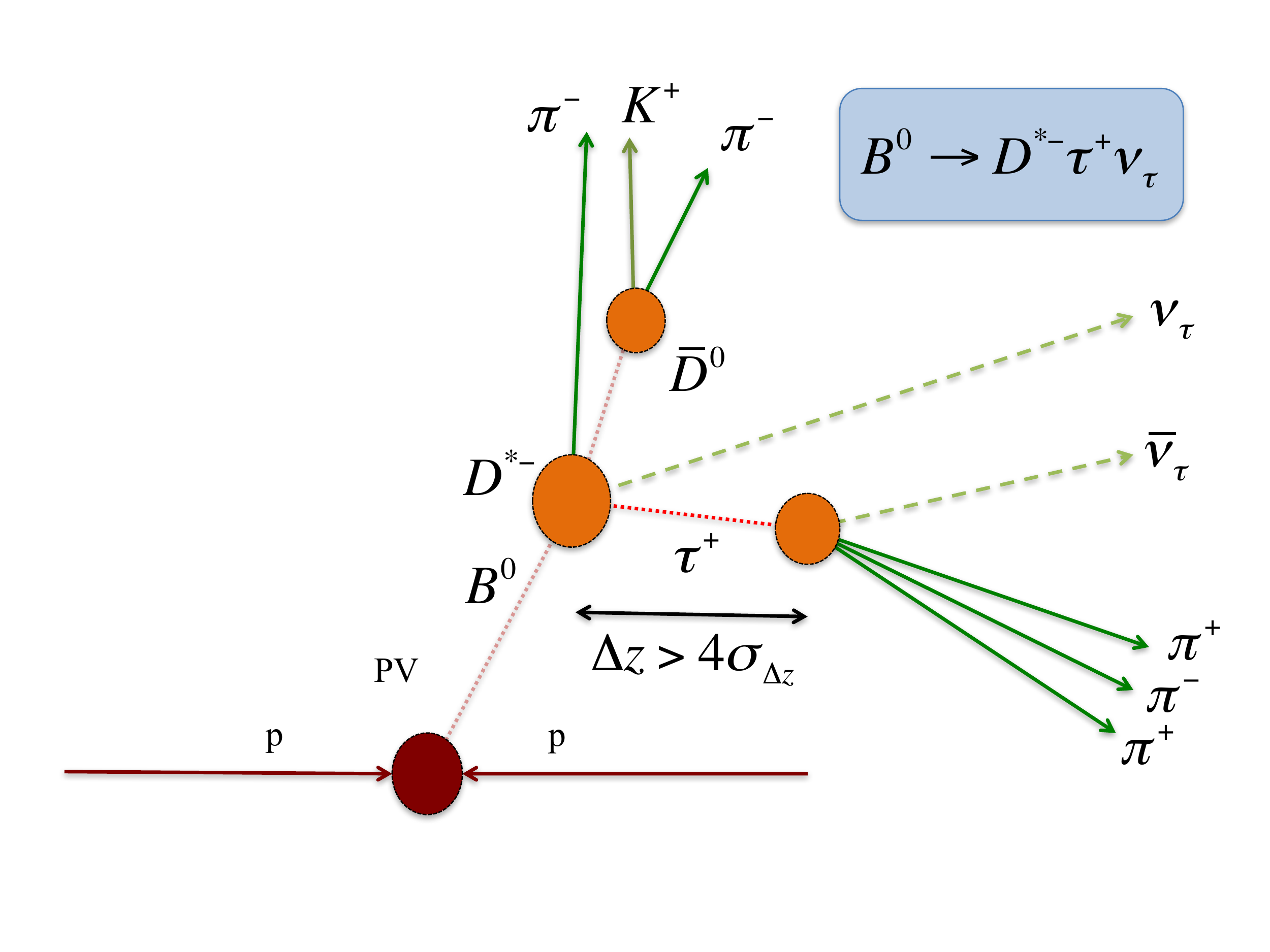}
	\caption{Topology of the signal decay. A requirement on the
          distance between the
          3\pion and the \Bz vertices along the beam direction to be greater than
          four times its uncertainty is applied. For $\PB\to\Dstar3\pion(X)$ decays, the 3\pion vertex coincides with the \PB vertex.}
\label{fig:sigtopo}
\end{figure}

One of the key aspects of this analysis is the necessary suppression 
of the large background originating from \bquark-hadron decays that include a \Dstarm meson, a 3\pion system, and any other
unreconstructed additional particles, $X$. 
This is achieved by
requiring that the position of the 3\pion vertex lies further away
from the proton-proton interaction vertex than that of the \Bz vertex,
as shown in Fig.~\ref{fig:sigtopo}. 
However, double-charm background processes due to \PB-meson decays into a \Dstarm and another
charmed hadron that subsequently decays into a final state containing
three charged pions, are topologically similar to the signal. %cannot be distinguished from signal in this way. 
The largest contribution originates from \decay{\B}{\Dstarm\Dsp(X)}
decays, where \PB  denotes a \Bz, \Bp or \Bs meson and the notation 
$(X)$ is used when unreconstructed particles may be present in the decay chain. Double-charm backgrounds are suppressed by means of a
multivariate algorithm~\cite{BBDT} which exploits the differences in the decay
dynamics and kinematics with respect to the signal process, together
with different properties used by partial reconstruction
algorithms. 
%
%CB TAKEN OUT AS PER REFEREES SUGGESTION
%The signal yield is measured using a three-dimensional fit
%to the invariant mass squared of the lepton pair ($q^2$), the decay
%time of the \tauon lepton ($t_{\tauon}$), and the output of the multivariate
%classifier. In this fit, template distributions for signal and
%background are determined from simulation.

The \lhcb detector is a 
single-arm forward spectrometer covering the pseudorapidity range 
$2 < \eta < 5$, described in detail in Refs.~\cite{Alves:2008zz,LHCb-DP-2014-002}.
In the simulation, proton-proton collisions are generated using
\pythia~\cite{Sjostrand:2006za,*Sjostrand:2007gs} 
with a specific \lhcb
configuration~\cite{LHCb-PROC-2010-056}.  Decays of hadronic particles
are described by \evtgen~\cite{Lange:2001uf}, in which final-state
radiation is generated using \photos~\cite{Golonka:2005pn}. 
The {\mbox{\textsc{Tauola}}\xspace} package~\cite{Davidson:2010rw} is used
to simulate the decays of the \tauon lepton into 3\pion\neutb and
3\pion\piz\neutb final states, according to the resonance chiral Lagrangian model~\cite{Nugent:2013hxa} with a tuning
based on the results from the \babar collaboration~\cite{Nugent:2013ij}. 
The interaction of the generated particles with the detector, and its response,
are implemented using the \geant
toolkit~\cite{Allison:2006ve, *Agostinelli:2002hh} as described in
Ref.~\cite{LHCb-PROC-2011-006}.
The signal decays are simulated using form factors that are derived
from the heavy-quark effective theory~\cite{Caprini:1997mu}. The experimental
values of the corresponding parameters are taken from Ref.~\cite{HFAG}, except for an
unmeasured helicity-suppressed component, which is taken 
from Ref.~\cite{Korner:1989qb}. 

The online event selection is performed by a
trigger system~\cite{LHCb-DP-2012-004}, 
which consists of a hardware stage based on information from the calorimeter and muon
systems, followed by a software stage that performs a full event
reconstruction.
At the hardware stage, events are selected if either particles forming
the signal candidate satisfy  a requirement on
    transverse energy, or
particles other than those forming the 
signal candidate pass any trigger algorithm. 

The software trigger requires a two-, three-, or four-track secondary
vertex with significant displacement from any primary proton-proton
interaction vertex (PV) consistent with the decay of a \bquark hadron, or
a two-track vertex with a significant displacement from any PV
consistent with a $\Dzb\to K^+\pi^-$ decay. In both cases, at least one
charged particle must have a transverse momentum $\pt > 1.7\gevc$ and
must be
inconsistent with originating from any PV. A multivariate
algorithm~\cite{BBDT}
is used for the identification of secondary vertices consistent with
the decay of a \bquark hadron, while secondary vertices consistent with the
decay of a \Dzb meson are identified using topological  
criteria.

In the offline selection, \Dzb, \Dstarm and \tauon candidates are selected based
on kinematic, geometric, and particle identification criteria. 
Three charged pions are used to reconstruct $\tau$-decay candidates,
including both the \decay{\taup}{3\pion\neutb} and
\decay{\taup}{3\pion\piz\neutb} modes. 
The vertex position and the momentum of the \Bz candidate are determined through a
 fit to all reconstructed particles in the  
decay chain~\cite{Hulsbergen:2005pu}. The difference of the positions of the 3\pion and the
\Bz vertices along the beam direction, divided by its uncertainty, has
to be greater than four. This requirement 
suppresses the background due to $B\to\Dstarm3\pion X$ decays by three
orders of magnitude and has an efficiency of 35\% for the signal.  
The normalization sample is selected by requiring the
difference in the positions of the \Dzb and 3\pion vertices along the
beam direction, divided by its uncertainty, to be greater than four. 

Backgrounds due to partially reconstructed \PB-meson decays, where at least
one additional particle originates from either the 3\pion vertex or the
\PB vertex, or from both, are suppressed by requiring a single \Bz candidate
per event. In addition, a charged-particle isolation algorithm is applied as described
in the following. 
Tracks other than those used for the signal candidate are considered
if they have minimal requirements on the transverse momentum and
are inconsistent with originating from any PV.
If any of these tracks has an impact parameter significance with respect to either
the \Bz or \tauon vertex smaller than 5 standard deviations, the \Bz candidate is
rejected. This criterion rejects 95\% of candidates due to
$\PB\to\Dstarm\Dz (X)$ decays, while retaining  80\% of the signal decays. 
In addition, a neutral-particle isolation algorithm
computes the multiplicities of reconstructed tracks and neutral particles, and the energy in the calorimeter system, 
contained in a cone centered around the direction of the \tauon candidates. 
These variables are used as inputs of the multivariate classifier
described below.  

Variables such as the squared invariant mass of the $(\tau, \nu_{\tau})$ 
pair, $q^2$ , and the $\tau$ decay time, $t_{\tau}$, provide good discrimination 
between signal and background processes, but they depend on the momenta
of the neutrinos in the final state of the \Bz decay. 
However, due to the presence of a single neutrino in the 
\tauon decay, the momentum of the \tauon lepton 
can be determined, up to a two-fold ambiguity, from the momentum
vector of the 3\pion system and the flight direction of the \tauon 
candidate. 
The value of the \tauon momentum is approximated by taking the average
of the two solutions, as discussed in Ref.~\cite{ourPRD}. 
A similar strategy is used to compute the \Bz momentum. 
The \Bz rest frame variables are determined with sufficient accuracy
to retain their discriminating power. 
A partial reconstruction is performed also under the background hypothesis
where $\Bz\to\Dstarm\Dsp (\to 3\pi N)$, with $N$ denoting a neutral
system.   The variables describing decay kinematics, as reconstructed
by this algorithm, differ between signal and background processes; a
selected set is used as input to the multivariate classifier described
below. 

The dominant double-charm background process
\decay{\B}{\Dstarm\Dsp(X)} is reduced by taking into account the resonant 
structure of the 3\pion system. 
The \taup lepton decays to 3\pion final states predominantly through the
$a_1(1260)^+\to\rhoz\pip$ decay. 
By contrast, the \Dsp meson decays to 3\pion final states 
predominantly through the $\eta$ and $\eta^\prime$ resonances. 
These and other features are exploited by means of a 
boosted decision tree
(BDT) ~\cite{Breiman,AdaBoost}, as described in Ref.~\cite{ourPRD}. 
The BDT response in the simulation is validated 
using three control samples: a $\PB\to\Dstarm\Dsp (X)$ data sample
which is obtained by using partial reconstruction under the background
hypothesis; 
a $\PB\to\Dstarm\Dz (X)$ data sample, with the subsequent 
\decay{\Dz}{K^-3\pion} decay, which is obtained by 
removing the charged-particle isolation criterion and requiring a
particle satisfying kaon identification criteria with an origin at the 3\pion vertex; and a $\PB\to\Dstarm\Dp
(X)$ data sample, with \Dp\to\Km\pip\pip, which is 
obtained replacing the negative pion with a candidate 
identified as a kaon. For all these samples, good
agreement between data and simulation is
observed in the distributions of the variables used in the BDT. These control samples are also
used to correct the simulation to reproduce the expected 
distributions of the fit variables in data. 

The yield of the normalization mode is determined by fitting the 
invariant mass distribution of the $\Dstarm3\pion$ system around the
known \Bz mass~\cite{PDG2017} for candidates in the normalization sample. The fitting
function of the normalization channel is the sum of
a Gaussian function and a Crystal Ball
function~\cite{Skwarnicki:1986xj}. 
An exponential
function is used for the combinatorial background. All parameters are
floating in the fit. A total of $N_{\mathrm{norm}} = 17\,660 \pm 158$ 
candidates are found,
where a small contribution of $151 \pm 22$ $B^0\to\Dstarm\Dsp (\to 3\pi)$ 
decays has been accounted for in the yield and uncertainty.
The latter component is estimated by fitting the 3\pion mass
distribution for candidates with a reconstructed \Bz mass in a window around the
known value. 

The signal yield is obtained from a three-dimensional binned fit to the data,
in a region of the BDT output enriched in signal decays. The fit
dimensions are $q^2$, $t_{\tau}$ and the BDT output.  Several 
components enter in the fit. In particular, a signal component which
also accounts for 
higher-mass charm-meson states; background components due to $\B\to\Dstarm\Dsp (X)$,
$\B\to\Dstarm\Dp (X)$ and $\B\to\Dstarm\Dz (X)$ decays; a residual
contribution from $\PB\to\Dstarm3\pion X$ decays; and a combinatorial background. 

The signal template is the sum of two terms, due to $\taup\to 3\pi\neutb$ and
$\taup\to 3\pi\piz\neutb$ decays, where the relative ratio between
these components is 
fixed according to their 
branching fractions and simulation-derived selection efficiencies. 
A contribution due to $\PB\to D^{**}\taup\neut$ decays, where $D^{**}$
denotes $P$-wave charm mesons or any higher mass states, with the \Dstarm
being produced in the $D^{**}$ decay chain, is also related to the
signal yield through a proportionality factor derived from
Ref.~\cite{Scora:1995ty}. 
A data sample where the narrow $D^0_1(2420)$
and $D^{*0}_2(2460)$ resonances are reconstructed in their $\Dstar\pi$
decays is used to validate the simulation. 

\begin{figure}[!tb]
  \begin{center}
    \includegraphics[width=0.68\textwidth]{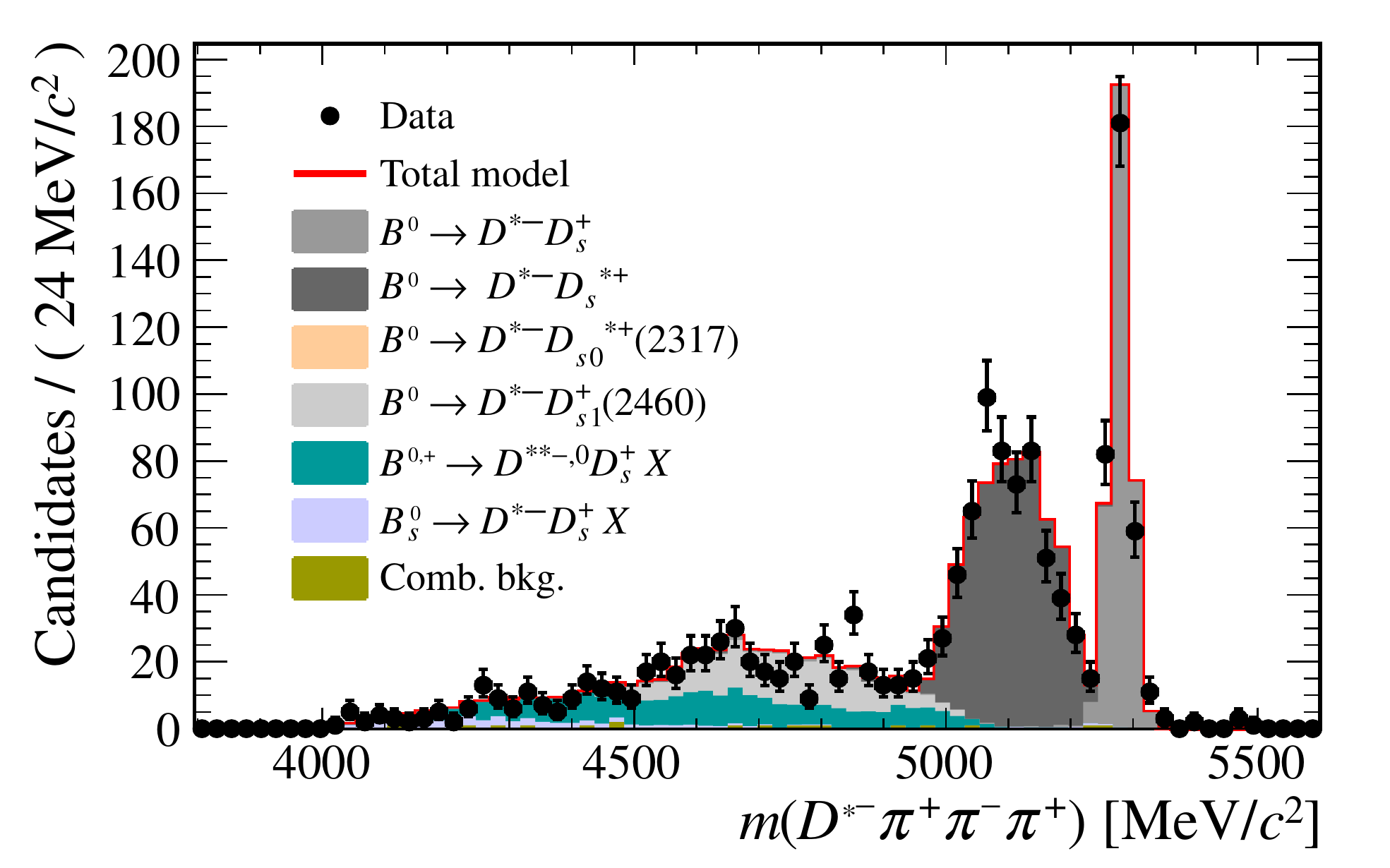}
  \end{center}
  \caption{
    \small %captions should be a little bit smaller than main text
{Results from the fit to the invariant mass of the \Dstarm\Dsp pair for the 
$\Dstarm\Dsp (X)$ data control sample,
with $\Dsp\to 3\pi$. 
The components contributing to the fit model are indicated in the 
legend. }
   }
  \label{fig:Ds_control_Fit}
\end{figure}

The background originating from decays of \PB mesons into $\Dstarm\Dsp
(X)$ final states is divided
into contributions from $\Bz\to\Dstarm\Dsp$,
$\Bz\to\Dstarm\Dssp$, $\Bz\to\Dstarm D^{*+}_{s0}(2317)$, $\Bz\to\Dstarm
D^{+}_{s1}(2460)$, $\B\to D^{**}\Dsp X$ and $B^0_s\to\Dstarm\Dsp
X$. The relative yield of each of these processes is constrained in the final fit using the results
of an auxiliary fit, shown in Fig.~\ref{fig:Ds_control_Fit}, 
to the $\Dstarm3\pion$ invariant mass. The fit is performed on a 
control sample of data obtained
by reconstructing the $\Dsp\to\pip\pim\pip$ decay. 

\begin{figure}[!tb]
 \begin{center}
   \includegraphics*[width=0.68\textwidth]{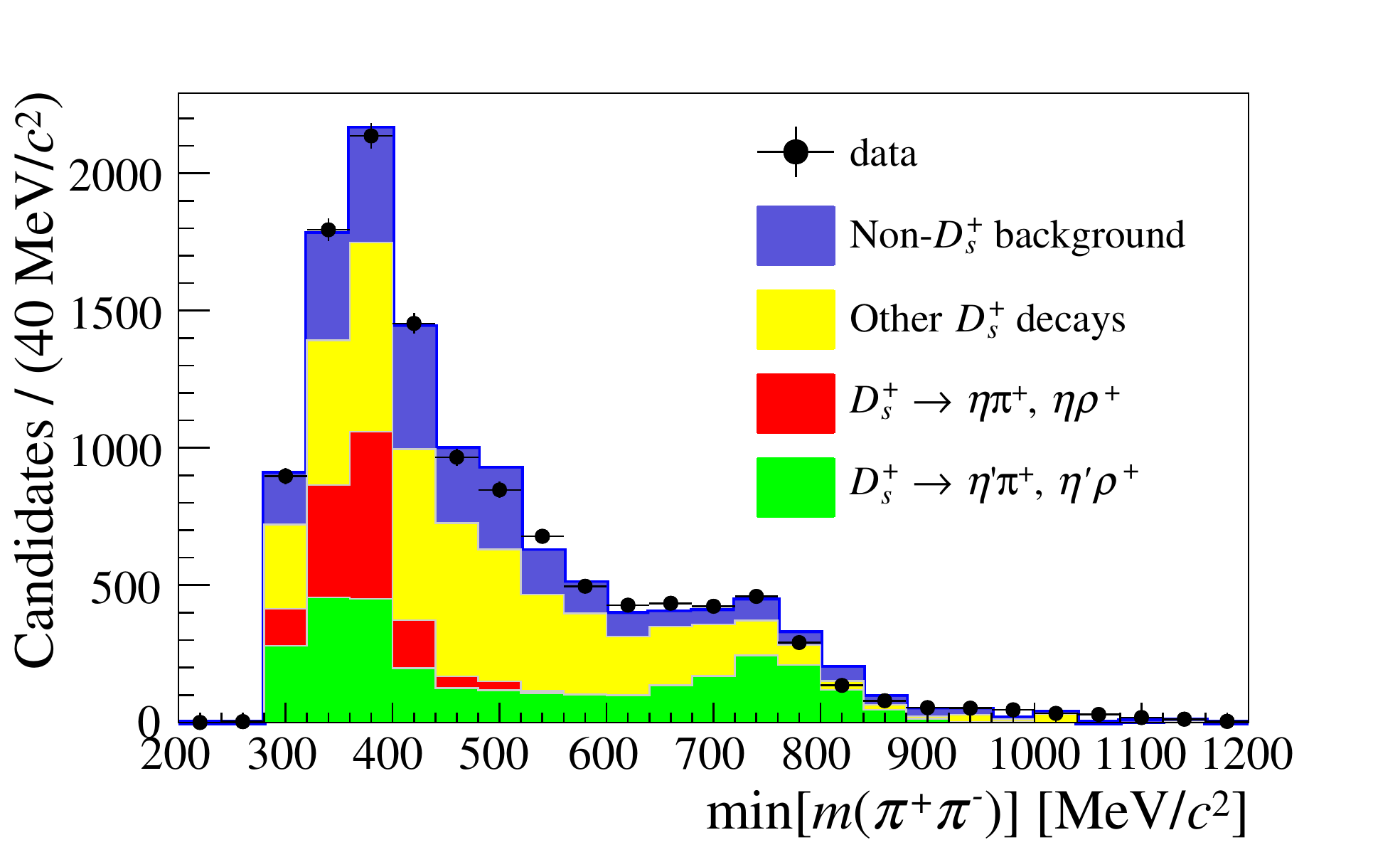}
 \end{center}
\caption{Distribution of ${\mathrm{min}}[m(\pi^+\pi^-)]$ 
  for a sample enriched in \decay{\PB}{\Dstarm\Dsp(X)} decays, obtained by
requiring the BDT output below a threshold. The different fit
components are indicated in the legend.}
 \label{fig:dsmodelfit}
\end{figure}

The \Dsp decay model used in the simulation does not accurately describe the
data because of the limited knowledge of the \Dsp decay amplitude
to 3$\pi$$X$ final states.
Therefore, the contribution of the background from \Dsp decays is
determined from data in a control region, selected by the BDT
output, where this background is abundant. In this region, 
the distributions of the minimum and maximum invariant masses of the
oppositely charged pions, ${\mathrm{min}}[m(\pip\pim)]$ and ${\mathrm{max}}[m(\pip\pim)]$, the invariant mass of the same-charge
pion pair and that of the 3\pion system are fitted simultaneously in
order to determine the contributions from different \Dsp final states. 
These are grouped in four
categories. The first (second) includes \Dsp decays into $\eta\pi$ or 
$\eta\rho$ ($\eta^{\prime}\pi$ or 
$\eta^{\prime}\rho$), where at least one pion originates from the $\eta$
($\eta^{\prime}$) decay. The third category contains \Dsp decays where at least one pion originates from another intermediate resonance such as
an $\omega$ or $\phi$ meson, \decay{\Dsp}{3\pion X} decays 
where none of the three pions originates from an intermediate
resonance, and \decay{\Dsp}{\taup(\to 3\pion\neutb)\neut} decays. The fourth category
consists of backgrounds without \Dsp mesons. 
Figure~\ref{fig:dsmodelfit} shows, as an example, the distribution of
${\mathrm{min}}[m(\pi^+\pi^-)]$ and the resulting fit components. 
The results obtained by the fit in this region of BDT output are used
to compute weights for each \Dsp decay mode, to be applied to the 
simulation. The templates used for these decays in the BDT output
region considered in the final fit are then recomputed by taking from
simulation the relative proportion between the yields in the two
regions of the BDT
output for each decay mode. 

Background originating from $\B\to\Dstarm\Dz X$ decays is subdivided into two contributions,
depending on whether the 3\pion system originates from the same \Dz vertex, or
whether one pion originates from the \Dz vertex and the other
two from elsewhere. The contribution of the former background is constrained by 
the yield obtained from the $\B\to\Dstarm\Dz (X)$ control
sample. 
The template shape is also
validated using this control sample. The yield of the other
$\B\to\Dstarm\Dz X$ background component is a free
parameter in the fit, while its shape is taken from simulation. 
The yield of the $\B\to\Dstarm\Dp X$ background is also a free
parameter. The template shape is validated using the corresponding control
sample. 
A residual background from $\PB\to\Dstarm3\pion X$ modes is included in the fit.
The yields of these components are constrained by those measured from
a data sample enriched with \decay{\PB}{\Dstarm 3\pion X} decays in which the distance of the
\PB vertex from the PV exceeds that of the 3\pion.

The combinatorial background is divided into two contributions, depending
on whether the background contains a real \decay{\Dstarm}{\Dzb\pim} decay chain or not. In the first case, the
\Dstarm and the 3\pion systems are required to originate from different \PB\ decays. The
templates for this background are taken from simulation. A sample
of candidates where the \Dstarm and the 3\pion systems have the same
charge is used to normalize data and simulation in the region where
the $\Dstarm3\pion$ mass is above the known \PB mass. The background not
including a real \Dstarm decay chain is parameterized and constrained 
using candidates outside a window around the known \Dzb mass. 

\begin{figure}[!htbp]
   \begin{center}
     \includegraphics[width=0.9\textwidth]{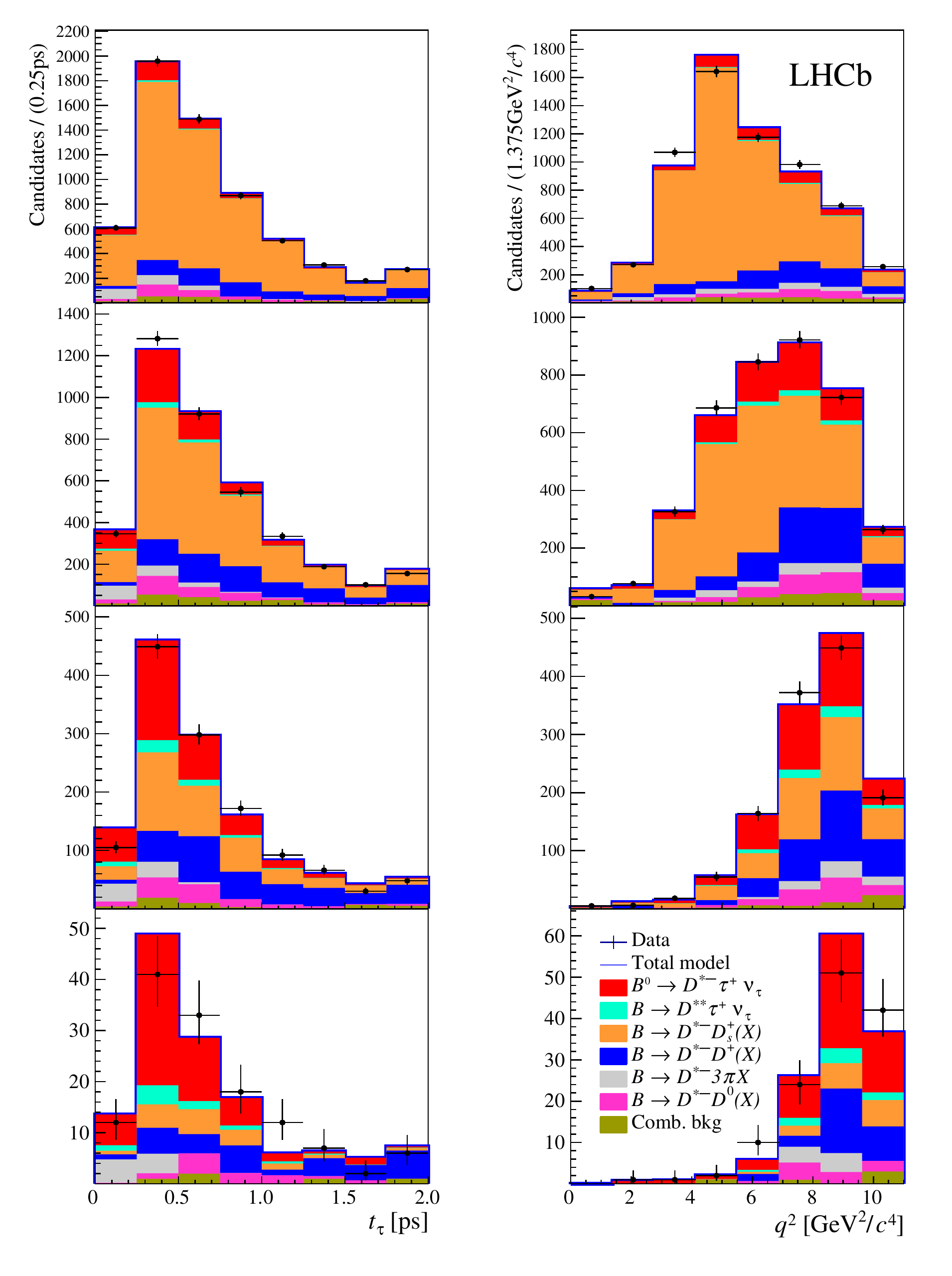}
    \end{center}
   \caption{
     \small %captions should be a little bit smaller than main text
   Distributions of (left) $t_{\tau}$ and (right) $q^2$ in
   four different BDT bins, with increasing values of the BDT response  
 from top to bottom.  The
 various fit components are described in the legend. 
  }
   \label{fig:fit_results_nominal_histfact_BDT}
 \end{figure}

The results of the fit are shown in
Fig.~\ref{fig:fit_results_nominal_histfact_BDT}. The global \chisq of
the fit is 1.15 per degree of freedom, after taking into
 account the statistical fluctuation in the simulation templates. 
The signal 
yield is corrected for a small bias of $40$~candidates, due to the finite size of the
templates from simulation, as detailed below, giving
$N_{\mathrm{sig}}=1296 \pm 86$ 
candidates. The result  
\begin{equation*}
{\cal{K}}(\Dstarm) =
1.97 \pm 0.13\stat \pm 0.18\syst
\end{equation*}
\noindent is determined from Eq.~\ref{eqn:kappa}, where
    the  efficiencies for events within LHCb acceptance are  ($0.39 \times 10^{-3}$) and
     ($1.36 \times 10^{-3}$) for signal and normalization modes, respectively, are taken from
    simulation, and an effective sum $(13.81\pm 0.07)\%$ of the branching fractions for the $\taup\to 3\pi\neutb$ and $\taup\to
3\pi\piz\neutb$ decays is used to account for the different selection
efficiencies between the two modes and small feeddown from other
\tauon decays. A 
correction factor 
$1.056 \pm 0.025$ has also been applied 
to account for discrepancies between data and simulation, 
and for a small feeddown contribution from
{\decay{\Bs}{D_s^{**-}\taup\neut}} decays, where
{\decay{D_s^{**-}}{\Dstarm K^0}}. 

The branching fraction 
\begin{equation*}
{\cal{B}}(\Bz\to\Dstarm\taup\neut)
=\left(1.42\pm 0.094\stat \pm0.129\syst
\pm0.054\extrn\right)\times 10^{-2}
\end{equation*}
\noindent is obtained by using ${\cal{B}}(\Bz\to\Dstarm 3\pi)=(7.214
\pm 0.28) \times 10^{-3}$, the weighted average of the
\lhcb~\cite{LHCb-PAPER-2012-046}, \babar~\cite{TheBABAR:2016vzj}, and \belle~\cite{Majumder:2004su} measurements.
Finally, the ratio of branching fractions 
\begin{equation*}
{\cal{R}}(\Dstarm) = 0.291 \pm 0.019 \stat
    \pm 0.026 \syst \pm 0.013 \extrn 
\end{equation*}
is obtained by using ${\cal{B}}(\Bz\to\Dstarm\mup\neum) = (4.88\pm 0.10)
\times 10^{-2}$ from Ref.~\cite{HFAG}. 
\noindent In both results, the 
third uncertainty is due to the limited knowledge of the
external branching fractions. 

Systematic uncertainties on ${\cal{R}}(\Dstarm)$ are reported in Table~\ref{tab:systematics}. 
\begin{table}
  \centering
  \caption{Relative systematic uncertainties on ${\cal{R}}(\Dstarm)$.}
  \label{tab:systematics}
    \begin{tabular}{l|c}
      \hline
      Source                          & $\delta R(\Dstarm) /
                                        R(\Dstarm) [\%]$ \\
      \hline
      Simulated sample size     & 4.7 \\ % 4.1 in quadrature with 1.7 and 1.6
      Empty bins in templates                            & $1.3$ \\
      Signal decay model               &  1.8 \\ % 0.7 in quadrature with 1.0, 0.4, 1 and 0.7 
      $D^{**}$\tauon\neu and $D_s^{**}$\tauon\neu feeddowns & $2.7$ \\ %1.5 in quadrature with 2.3
      $D_s^+ \to 3\pi X$ decay model         & $2.5$ \\
      $B\to D^{*-} D_s^+ X$, $B\to D^{*-} D^+ X$, $B\to D^{*-} D^0 X$ backgrounds & 3.9 \\ % 2.9 in quadrature with 2.6
      Combinatorial background              & $0.7$ \\
      $\PB\to\Dstarm3\pion X$ background & 2.8 \\ % 1.9 in quadrature with 2.1
      Efficiency ratio                  & 3.9 \\ %quadrature of the above 8
      Normalization channel efficiency (modeling of $\Bz\to\Dstarm 3\pi$) & $2.0$ \\ 
      \hline
      Total uncertainty                  & $9.1$ \\
      \hline
    \end{tabular}
\end{table}
The uncertainty 
due to the limited size of the simulated
samples is computed by repeatedly sampling each template with a bootstrap
procedure, performing the fit, and taking the standard deviation of the
results obtained. 
Empty bins in the
templates used in the fit also introduce a positive bias of 3\% in the determination of the signal
yield. This corresponds to a correction of 40 candidates, with an
uncertainty of 1.3\%. The
limited size of the simulated samples also contributes
to the systematic uncertainty 
on the efficiencies for signal and
normalization modes. 

The systematic uncertainty associated with the signal decay model
derives from the limited knowledge of the form
factors and the $\tau$ polarization, from possible contributions from other
$\tau$ decay modes, and from the relative branching fractions and
selection efficiencies of $\taup\to 3\pi\piz\neutb$ and 
$\taup\to 3\pi\neutb$ decays. 
Uncertainties due to knowledge of the $D^{**}$\taup\neut
contribution to the signal yield are estimated using a control
sample where one additional charged pion originating from the \B
vertex is identified. The observed yield of the narrow
$D_1(2420)^0$ resonance is used to infer a 40\% uncertainty on the
yield of $D^{**}$\taup\neut decays relative to that of the signal.
A systematic uncertainty is also assigned to take into account the feeddown from \Bs decays into $D_s^{**-}$\taup\neut. 

The uncertainty due to the knowledge of the \Dsp decay model is 
estimated by repeatedly varying the correction factors of the
templates within their uncertainties, as determined from the associated
control sample, and performing the fit. 
The spread of the fit results 
is assigned as the
corresponding systematic uncertainty.
The template shapes of the \Dstarm\Dsp, \Dstarm\Dz and
\Dstarm\Dp backgrounds depend on the dynamics of the
corresponding decays. Empirical variations of the kinematic
distribution are performed, and the spread of the fit results is 
taken as a systematic uncertainty. A similar procedure is applied to the template for the
combinatorial background. 
Other sources of systematic uncertainty arise from
the inaccuracy on the yields of the various background
contributions, 
and from the limited knowledge of
the normalization modeling and the resonant structure of the residual background
due to $\B\to\Dstarm3\pion X$ decays. 

Systematic effects on the efficiencies for signal and normalization
partially cancel in the ratio. 
The trigger efficiency depends on the distributions of the decay time of the 3\pion system and 
the invariant mass of the $\Dstarm 3\pion$ system. These distributions
differ between the signal and normalization modes, and the difference of the trigger efficiency for these two decays is taken into account.

In conclusion, the first measurement of ${\cal{R}}(\Dstarm)$ with
three-prong \tauon decays has been performed by using a 
technique that is complementary to all previous measurements of this
quantity and offers the possibility to study other \bquark-hadron decay modes in a
similar way. The result,
${\cal{R}}(D^{*-}) = 0.291 \pm 0.019 \stat
    \pm 0.026 \syst \pm 0.013 \extrn$, is one of the most precise
    single  measurements performed so far. It is 1.1 
   standard deviations higher than the SM calculation 
   ($0.252 \pm 0.003$) of Ref.~\cite{Fajfer:2012vx}, and consistent
   with previous determinations. 
An average
of this measurement with the \lhcb result using
\decay{\taup}{\mup\neum\neutb} decays~\cite{LHCb-PAPER-2015-025}, accounting for small
correlations due to form factors, \tauon polarization and $D^{**}\taup\neut$ feeddown, gives
${\cal{R}}(\Dstarm) = 0.31 \pm 0.016 \stat \pm 0.021 \syst$, consistent
with the world average and 2.2 standard deviations above the SM
prediction. 
The overall status of ${\cal{R}}(\D)$ and ${\cal{R}}(\Dstar)$
measurements is reported in Ref.~\cite{HFAG}. 
After the inclusion of this result, the combined discrepancy of
${\cal{R}}(\D)$ and ${\cal{R}}(\Dstar)$ determinations with the SM
prediction is 4.1 standard deviations.

\vspace{1cm}
\noindent We express our gratitude to our colleagues in the CERN
accelerator departments for the excellent performance of the LHC. We
thank the technical and administrative staff at the LHCb
institutes. We acknowledge support from CERN and from the national
agencies: CAPES, CNPq, FAPERJ and FINEP (Brazil); MOST and NSFC
(China); CNRS/IN2P3 (France); BMBF, DFG and MPG (Germany); INFN
(Italy); NWO (The Netherlands); MNiSW and NCN (Poland); MEN/IFA
(Romania); MinES and FASO (Russia); MinECo (Spain); SNSF and SER
(Switzerland); NASU (Ukraine); STFC (United Kingdom); NSF (USA).  We
acknowledge the computing resources that are provided by CERN, IN2P3
(France), KIT and DESY (Germany), INFN (Italy), SURF (The
Netherlands), PIC (Spain), GridPP (United Kingdom), RRCKI and Yandex
LLC (Russia), CSCS (Switzerland), IFIN-HH (Romania), CBPF (Brazil),
PL-GRID (Poland) and OSC (USA). We are indebted to the communities
behind the multiple open-source software packages on which we depend.
Individual groups or members have received support from AvH Foundation
(Germany), EPLANET, Marie Sk\l{}odowska-Curie Actions and ERC
(European Union), ANR, Labex P2IO, ENIGMASS and OCEVU, and R\'{e}gion
Auvergne-Rh\^{o}ne-Alpes (France), RFBR and Yandex LLC (Russia), GVA,
XuntaGal and GENCAT (Spain), Herchel Smith Fund, the Royal Society,
the English-Speaking Union and the Leverhulme Trust (United Kingdom).

%\input{appendix}
%
% This should be taken out in the final paper
%\input{supplementary-app}

\addcontentsline{toc}{section}{References}
\setboolean{inbibliography}{true}
\bibliographystyle{LHCb}
\bibliography{main,LHCb-PAPER,LHCb-CONF,LHCb-DP,LHCb-TDR}

 \newpage

% Author List ----------------------------                                                                                                                                                                                                                                                                                                
%  You need to get a new author list!                                                                                                                                                                                                                                                                                                    

\centerline{\large\bf LHCb collaboration}
\begin{flushleft}
\small
R.~Aaij$^{40}$,
B.~Adeva$^{39}$,
M.~Adinolfi$^{48}$,
Z.~Ajaltouni$^{5}$,
S.~Akar$^{59}$,
J.~Albrecht$^{10}$,
F.~Alessio$^{40}$,
M.~Alexander$^{53}$,
A.~Alfonso~Albero$^{38}$,
S.~Ali$^{43}$
% ...

{\footnotesize \it
% ...
$ ^{5}$Clermont Universit{\'e}, Universit{\'e} Blaise Pascal, CNRS/IN2P3, LPC, Clermont-Ferrand, France\\
% ...
$ ^{10}$Fakult{\"a}t Physik, Technische Universit{\"a}t Dortmund, Dortmund, Germany\\
% ...
$ ^{38}$ICCUB, Universitat de Barcelona, Barcelona, Spain\\
$ ^{39}$Universidad de Santiago de Compostela, Santiago de Compostela, Spain\\
$ ^{40}$European Organization for Nuclear Research (CERN), Geneva, Switzerland\\
% ...
$ ^{43}$Nikhef National Institute for Subatomic Physics, Amsterdam, The Netherlands\\
% ...
$ ^{48}$H.H. Wills Physics Laboratory, University of Bristol, Bristol, United Kingdom\\
% ...
$ ^{53}$School of Physics and Astronomy, University of Glasgow, Glasgow, United Kingdom\\
% ...
$ ^{59}$University of Cincinnati, Cincinnati, OH, United States\\
% ...
}
\end{flushleft}
 
%\newpage
%\input{LHCb_authorlist.tex}

\end{document}